\documentclass[useAMS,usenatbib,onecolumn]{mnras}
\usepackage{graphicx,rotating}
\usepackage{longtable}

 \title[''Scraggy'' dark halos around bulge-less spiral galaxies]{''Scraggy'' dark halos around bulge-less spiral galaxies} 

 \author[I. D. Karachentsev,   V. E. Karachentseva]{I. D. Karachentsev$^1$\thanks{E-mail: ikar@sao.ru}, V. E. Karachentseva$^2$ \\
$^1$Special Astrophysical Observatory, Russian Academy of Sciences, N.Arkhyz, 369167 Russia\\ 
$^2$Main Astronomical Observatory of National Academy of Sciences of Ukraine, Kiev, 03143 Ukraine} 

\begin{document}
\date{Accepted XXX. Received XXX; in original form XXX}

\pagerange{\pageref{firstpage}--\pageref{lastpage}} \pubyear{XXX}

\maketitle

\label{firstpage}
\begin{abstract}

 We use a sample of 220 face-on bulge-less galaxies situated in the low density environment to estimate their total mass via orbital motions of supposed rare satellites. Our inspection reveals 43 dwarf companions having the mean projected separation of 130 kpc and the mean-square velocity difference of 96 km/s. For them, we obtain the mean orbital-mass-to-K-band luminosity ratio of $20\pm3$. Seven bulge-less spirals in the Local Volume are also characterized by the low mean ratio, $M_{orb}/L_K = 22\pm5$. We conclude that bulge-less Sc-Scd-Sd galaxies have poor dark halos, about two times lower than that of bulgy spiral galaxies of the same stellar mass.
\end{abstract}

\begin{keywords}
galaxies: bulges -- galaxies: spiral -- galaxies
\end{keywords}

\section{Introduction.}
    Among the galaxies with stellar masses of more than, $10^{10}M_{\odot}$ the  flat disc-like objects without the evidence of the central bulge constitute about 10\%. Almost all of them are classified as late-type spirals: Sc, Scd, and Sd according to de Vaucouleurs et al. (1976). Some of them demonstrate small pseudo-bulges formed during the process of the secular disc evolution.  Unlike classical bulges, the pseudo-bulges show a nearly exponent stellar density profile and the evidences of current star formation (Kormendy \& Kennicutt 2004). The observed abundance of the spiral discs without spheroidal stellar component imposes a challenge to models of the hierarchical clustering of galaxies via  their consecutive merging (Kormendy et al. 2010).
     
     Bulge-less galaxies are the most distinguishable when they are seen strictly edge-on.  A Reference Flat Galaxy Catalog = RFGC (Karachentsev et al. 1999) lists 4236 ''flat'' galaxies covering the all sky. Karachentseva et al. (2016) have selected from the RFGC catalog 817 ultra-flat galaxies (UFG) with an apparent axial ratio  $a/b>10$. The environment of the UFGs is characterized by low spatial density, a deficit of massive satellites and almost total absence of dwarf spheroidal satellites with old  stellar population.
    
    Inner structure of the UFGs is practically unseen due to the disc inclination. To determine the bulge-less galaxy structural features, Karachentsev \& Karachentseva (2019) collected 220 galaxies of the Sc, Scd, and Sd types, seen almost face-on. According to statistical analysis, about half of the galaxy sample have bar-like substructures,  in most part of the discs the unresolved (star-like) nuclei are seen, and a considerable part  (27--50)\% of the face-on galaxies have peripheral distortion of their spiral pattern.
    
    There is an idea that secular stability of very thin stellar discs requires the presence a very massive dark halo  around them (Banerjee \& Jog 2013). The shape of this halos is supposed to be nearly spherically symmetric. Unfortunately, the rotation curves obtained for the UFGs, extend not so far from the center  (Uson \& Matthews 2003,  Makarov et al. 2001),  which makes it impossible to evaluate the true  dimension and mass of the dark halo. Determining the total dark halo mass by  radial velocities and projected separations of the satellites is more promising method. However, the difficulties arise even with such approach. The search for satellites with measured radial velocities around the UFGs shows that $\sim$60\% of ultra-flat galaxies have not any detected physical satellites inside the virial radius $R_{vir}\simeq 250$ kpc, about 30\% of the UFGs  enter together with a rather bright neighbors in scattered associations (filaments, walls), and only $\sim$10\% of ultra-flat galaxies are the dominant objects in physical multiple systems (Karachentsev et al. 2016). In the following we present the results of searches for satellites around 220 late type spiral galaxies seen face-on as well as the estimates of the total mass of the bulge-less galaxies.
\begin{table}
\caption{Face-on bulge-less galaxies with orbital mass estimates.}
\begin{tabular}{lcrcrcrcrr}\hline
Galaxy     &    $B_t$  &    $V_{LG}$&   mod D  &   T  &   $\log L_K$ & $\Delta V$ &  $R_p$&  $\log M_{orb}$&  $M_{orb}/L_K$\\
\hline
            &   mag    &    km/s   &   mag    &    &  $ L_{\odot}$      &  km/s  &    kpc  &  $M_{\odot}$      & \\
    (1)      &  (2)     &   (3)    &  (4)    &(5)  & (6) &      (7)   &   (8)    &   (9)  &     (10)\\
\hline
IC1562        &13.60      &3771     &33.56     &c     &10.69      &122          &60      &12.02    &21\\
NGC0255       &12.41      &1694     &31.70     &c     &10.45       &37         &173      &11.45    &10\\
               &          &         &           &     &           &-72         &243      &12.17    &52\\
ESO542-004    &15.00      &5657     &34.50     &d     &10.29      &-66         &199      &12.01    &52\\
UGC02043      &14.96      &5316     &34.40     &c     &10.67      &-18         &128      &10.69     &1\\
ESO479-022    &15.39      &7266     &35.07     &c     &10.61       &45         &168      &11.60    &10\\
UGC02323      &15.66      &8139     &35.33     &c     &10.99     &-281          &41      &12.58    &39\\
UGC02692      &14.07      &6337     &34.77     &c     &11.09      &-42          &92      &11.28     &2\\
               &          &         &          &      &           &114          &233     &12.55    &29\\
NGC1376       &12.85      &4137     &33.82     &c     &11.14     &-115          &134     &12.32    &15\\
              &           &         &         &       &          &-100          &157     &12.27    &14\\
NGC1599       &14.10      &3947     &33.73    &c      &10.60      &100          &155     &12.26    &46\\
UGC03703      &15.35      &7260     &35.14    &c      &10.72      &-96         &189      &12.31    &39\\
UGC04380      &15.05      &7554     &35.24    &c      &10.86      &134         &208      &12.64    &60\\
UGC05169      &15.44      &7699     &35.29    &d      &10.45      &-81          &41      &11.50    &11\\
NGC2967       &12.28      &1679     &32.16    &c      &10.80      &101         &191      &12.36    &36\\
UGC05474      &14.89      &5847     &34.71    &cd     &10.54       &42          &80      &11.22     &5\\
              &           &         &         &       &          &-117         &154      &12.39    &71\\
              &           &         &         &       &          &-67         &161      &11.93    &24\\
UGC05483      &15.13      &6014     &34.77    &c      &10.62      &-21          &46      &10.38     &1\\
NGC3506       &13.16      &6252     &34.87    &c      &11.42        &1         &135       &8.20     &0\\
PGC034006     &14.13      &7543     &35.26    &c      &11.25      &168         &190       &12.80   &35\\
NGC3596       &11.79      &1062     &31.32    &cd     &10.42     &-101          &25       &11.48   &12\\
              &           &         &         &       &          &-60           &60      &11.41   &10\\
NGC3938       &10.87       &841     &30.87    &c      &11.03     &-179          &156      &12.77   &55\\
              &            &        &         &       &              &-82          &166      &12.12   &12\\
              &            &        &         &       &              &-191         &168       &12.86   &68\\
              &            &        &         &       &              &35          &223      &11.51    &3\\
              &            &        &         &       &              &27          &251      &11.33    &2\\
IC3271        &14.57      &7083       &35.13   &c      &10.97      &63         &103       &11.68    &5\\
NGC4653       &12.77      &2471       &33.08   &c      &10.85      &37          &68       &11.04    &2\\
              &           &           &        &       &             &13          &96       &10.45    &0 \\  
NGC5434       &13.94      &4587       &34.21   &c      &10.84     &115          &41       &11.81    &9\\
NGC5468       &12.95      &2734       &33.40   &c      &10.90      &68          &55       &11.48    &4\\
PGC058201     &15.69      &8562       &35.51   &c      &10.70      &43          &54       &11.07    &2\\
IC1221        &14.59       &5706       &34.63  &cd     &10.64      &75          &96       &11.80   &14\\
NGC6821       &13.62       &1680       &31.86  &d      &10.33     &-28         &127       &11.07    &6\\
NGC7137       &13.05       &1977       &32.16   &c     &10.57     &152          &77       &12.32   &56\\
NGC7495       &13.76       &5133       &34.28   &c     &11.06     &-39          &73       &11.12    &1\\
NGC7535       &14.28       &4884       &34.17  &cd     &10.63       &8          &47        &9.55    &0\\
              &            &           &       &       &             &-100         &174        &12.31   &48  \\  
              &            &           &       &       &             &-15         &182        &10.68    &1\\
ESO605-016    &13.23      &7999        &35.28   &c     &11.55     &-68         &142        &11.89    &2\\ \hline
\multicolumn{10}{l}{(1) galaxy name;  (2) apparent $B$-magnitude from HyperLEDA (Makarov et al. 2014);} \\
\multicolumn{10}{l}{(3) radial velocity with respect to the Local Group center;}\\
\multicolumn{10}{l}{(4,5) distance modulus and morphological type from HyperLEDA;}\\
\multicolumn{10}{l}{(6)  logarithm of K-band luminosity expressed in the solar units; }\\
\multicolumn{10}{l}{(7)  radial velocity difference of satellite relative to the central galaxy;}\\
\multicolumn{10}{l}{(8) a satellite projected separation;}\\ 
 \multicolumn{10}{l}{(9) logarithm of orbital mass; (10) orbital mass-to-luminosity ratio.}
\end{tabular}
\end{table}
    \section{Satellites around face-on bulge-less galaxies.}
    To reveal the physical satellites, we have used the option ''To Search for Nearby Objects'' proposed in the  NASA Extracalactic Database (=NED, http://ned.ipac.caltech.edu/). Around each bright enough galaxy from the list by Karachentsev \& Karachentseva (2019) with absolute magnitude $M_B  < -19\fm0$  we have searched for companions with radial velocity difference $\mid \Delta V \mid < 300$ km/s ranging in the projected separation  $R_p =250$ kpc. In the next stage, we  left only those conditionally isolated cases where other neighbors brighter than the central object were absent within $R_p =750$ kpc around the face-on galaxy under study. In the process, we excluded a lot of neighboring objects that turned out to be stars or  parts of a normal galaxy. Altogether, we detected 43 satellites around 30 face-on galaxies. The remaining galaxies from our list do not have satellites in the specified limits of
 $\mid \Delta V \mid$ and $R_p$, or they have as neighbors other bright galaxies violating the isolation condition. 
   
   In Table 1 data on the face-on bulge-less galaxies and their satellites are given. 
Before presenting the data we have to do some explanations. The total $B$-magnitude is corrected for Galactic extinction; the inner extinction for face-on galaxies we prove to be zero. The $K$- band magnitude is calculated via  
the corrected $B$- magnitude and morphological type  (Melnyk et al. 2017).  Calculating the satellite projected separation $R_p$, we use the galaxy distance $D=V_{LG}/H_0$ with the Hubble parameter $H_0 = 73$ km/s/Mpc. For nearby galaxies situated in the Local Volume we attracted  their individual distance moduli from
http://www.sao.ru/lv/lvgdb. The orbital (Keplerian) mass of a galaxy, $M_{orb} = (16/\pi G) \times \Delta V^2 \times R_p$, is estimated assuming a random orientation of satellite orbits with the mean orbital eccentricity of  $\langle e^2\rangle =1/2$ (Karachentsev \& Kudrya 2014), where  $G$ is the gravitation constant.  Notes to columns of Table 1 are given below the Table. 
  \begin{figure} 
\includegraphics[scale=0.8]{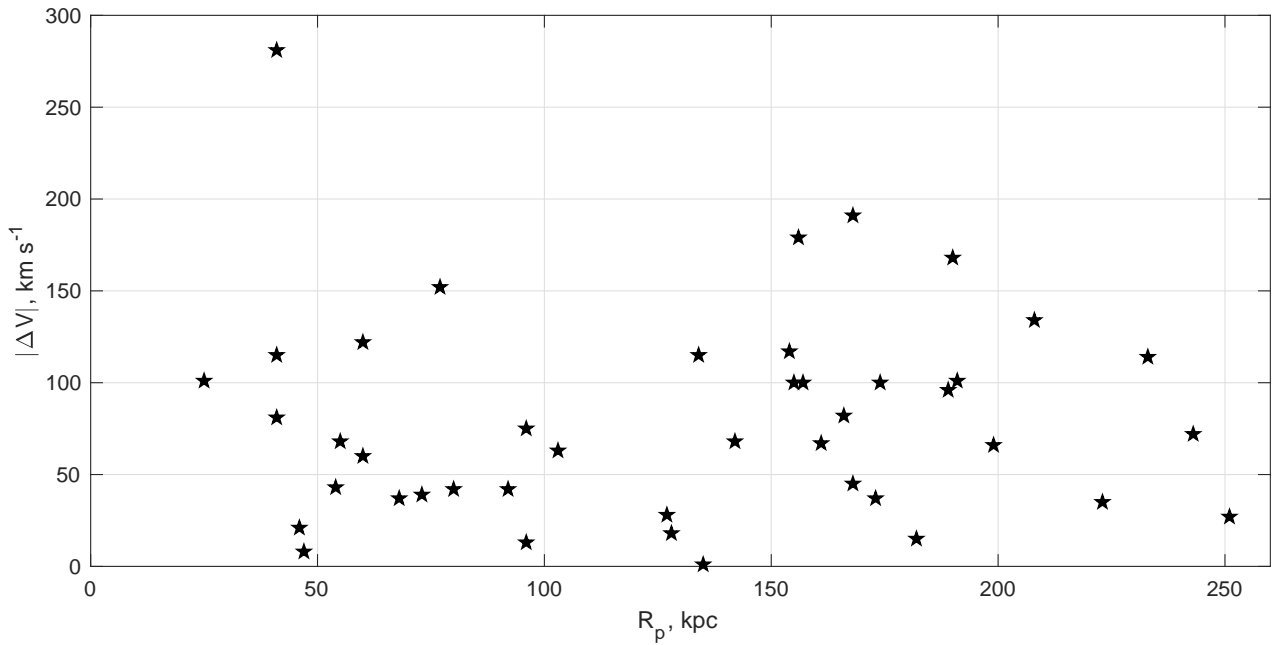} 
\caption{Distribution of the face-on bulge-less galaxy satellites on their radial velocity difference and projected separation.}
\end{figure} 
   The distribution of 43 satellites around their host face-on bulg-less galaxies on the modulus of radial velocity difference and the projected separation is presented in Fig.1. Here the mean-square of radial velocity difference is $\langle\Delta V^2\rangle^{1/2} = 96$ km/s, and the mean projected separation is equal to $\langle R_p\rangle =130$ kpc.
     
     As it is seen from Table 1, the majority of  face-on galaxies with detected satellites belong to the Sc morphological type. The median of their  luminosity, 10.71 dex, is comparable with the Milky Way luminosity. (This high value is caused by the fact that in searching for satellites we donated the face-on low luminosity galaxies with their shallow potential well.) The orbital mass values show a pronounced scatter that depends on prevailing character of satellites motions.  At fixed value of the orbital eccentricity, the ensemble mean of $M_{orb}$ is the unbiased estimate of the mass of central galaxy. In the case of an arbitrary eccentricity $e$, the orbital Keplerian mass estimate is expressed as 

$$M_{orb} = (32/3\pi)(1 - 2 e^2 /3)^{-1} \times G^{-1} \times \langle\Delta V^2  \times R_p\rangle.$$

This value grows in three times from severely round motions, $e = 0$, to the pure radial ones, $e = 1$. We assumed the mean value of   $\langle e^2\rangle =1/2$, corresponding the expected average obtained in the N-body simulations (Barber et al. 2014).
   
   The mean value of the orbital-mass-to-K-band luminosity ratio for face-on bulge-less galaxies is $\langle M_{orb} /L_K\rangle =20\pm3$ (see last column of Table 1). This ratio diminishes to $17\pm3$ having regard to statistical correction for errors of the radial velocity measurements.

    \section{Satellites of the face-on bulge-less galaxies in the Local Volume.}
      The Local Volume limited by the radius of 11 Mpc, is unique in having an abundance of galaxies which distances are measured at the Hubble Space Telescope with accuracy of about 5\%. The summary sample on the galaxy distances is presented in the Updated Nearby Galaxy Catalog = UNGC (Karachentsev et al. 2013) as well as in the regularly replenished data base http://www.sao.ru/lv/lvdb (Kaisina et al. 2012). In the Local Volume faint dwarf galaxies are seen which are lost in more distant and less researched regions of the universe.  The knowledge of the high accuracy measured distances permits to separate the physical satellites of bright galaxies  against foreground and background objects with more confidence. 
     In the Local Volume there are seven  face-on bulge-less galaxies from the list of Karachentsev \& Karachentseva (2019): IC~342 (3.28 Mpc), NGC~5068 (5.15 Mpc),  M~101  (6.95 Mpc),  NGC~6946  (7.73 Mpc),  NGC~3344  (9.82 Mpc),  NGC~628  (10.19 Mpc), and NGC~3184  (11.12 Mpc). Here in parentheses the distances from observer are given. Of them, two low luminosity galaxies, NGC~3344 and NGC~5068 have no satellites with measured radial velocities. The remaining  five galaxies have from 5 to 9 physical satellites, the data on which are presented in Table 2. Its columns contain: (1)  galaxy name, (2)  the total apparent $B$-magnitude, (3)  the projected separation in kpc, (4)  the difference between the radial velocities of the satellite and  the host galaxy, in km/s.     
    
\begin{table}
 \caption{Companions of face-on bulge-less galaxies in the Local Volume.}
 \begin{tabular}{lrrr}\hline

 Galaxy     &     $B_t$  &    $R_p$     &  $\Delta V$\\
\hline
	       &  mag  &  kpc  &  km/s\\
\hline
 {\bf IC 342}           &9.1       &0      &0\\
  &      &       &      \\  
 KK35           &15.7    &15     &-97\\
 UGCA 86     &13.5    &89      &36\\
 NGC 1560    &12.1   &312    &-67\\
 NGC 1569    &11.8   &312   &-126\\
 Cam A          &14.8   &326    &-77\\
 Cam B          &16.1   &365     &46\\
 Cas 1             &15.3   &526      &4\\
\hline
{\bf M 101}                 &8.3      &0         &0\\
  &      &       &      \\
 NGC 5477         &14.2    &44      &73\\
 PGC2448110     &17.3    &92     &14\\
 NGC 5474         &11.5    &96     &46\\
 UGC 8882        &15.8   &110    &104\\
 GBT1355+54       &-    &152     &-33\\
 Holm IV           &13.8   &163   &-106 \\  
 NGC 5585        &11.2   &429     &79\\
 dw1343+58       &15.7   &605    &-13\\
 DDO194           &14.5   &647      &3\\
\hline
 {\bf NGC6946}         &9.6     &0      &0\\
  &      &       &      \\
 KK 251           &16.5    &77     &86\\
 UGC 11583    &15.9    &86     &87\\
 KK 252          &16.7   &105     &99\\
 KKR 55          &17.0   &178     &-8\\
 KKR 56          &17.6   &311    &-83\\
 Cepheus 1      &15.4   &525     &18\\
 KKR 59          &15.7   &631    &-43 \\ 
 KKR 60          &18.0   &672    &-54\\
\hline
 {\bf NGC 628}           &9.8     &0      &0\\
  &      &       &      \\
 UGC 1171        &15.7   &132     &81\\
 UGC 1176        &14.4   &150    &-27\\
 SDSS0138+14 &18.0   &153     &86\\
 KDG 10           &16.3   &296    &132\\
 AGC 112454    &17.5   &297     &14\\
 UGC 1056       &14.8   &374    &-62\\
 JKB 142          &18.4   &404     &65\\
 UGC 1104        &14.5   &481     &29\\
 SDSS0142+13 &17.4   &501     &15\\
\hline
 {\bf NGC 3184}        &10.3     &0      &0\\
  &      &       &      \\
 KUG1013+414 &15.4    &90    &-85\\
 SDSS1028+42  &17.5   &445    &-34\\
 NGC 3104        &13.6   &543      &8\\
 SDSS1025+43  &17.3   &561     &58\\
 KUG1004+392  &15.6   &629      &9\\ \hline
 \end{tabular}
 \end{table}
\begin{figure}
\includegraphics[scale=0.8]{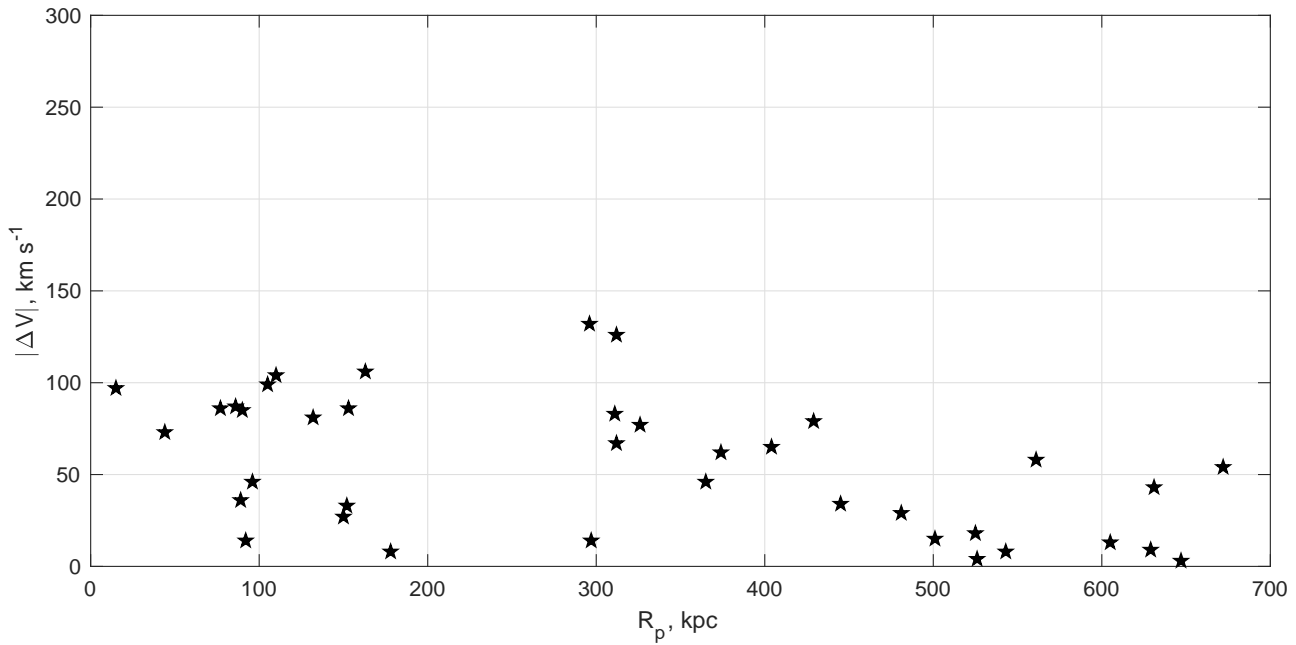} 
\caption{Distribution of satellites around the nearby face-on bulge-less galaxies: IC~342, M~101, NGC~6946, NGC~628, and NGC~3184 on their radial velocity difference and projected separation. }
\end{figure}
    The distribution of 38 satellites on their modulus velocity difference and projected separation is presented in Fig.2. As one can see, the satellite velocity differences do not exceed 150 km/s, justifying the above selection limit of $\mid \Delta V\mid < 300$ km/s in the search for galaxies situated in more distant volume. About half of satellites are located outside of a characteristic virial radius $\sim250$ kpc. Nethertheless, they all are inside the ''zero velocity sphere'' ($R_0\sim 1$ Mpc), which separates the  collapsing group members against the global cosmic expansion. Obviously, beyond the Local Volume, many distant satellites with $R_p > R_{vir}$  are lost among the general field galaxies.
   \begin{table}
\caption{Suites of companions around the Local Volume bulge-less spirals.}
\begin{tabular}{lrrrrrrrr}\hline
 Galaxy   &    D       &   n  &  $\Delta m_{12}$ & $\langle R_p\rangle$& $\sigma_V$ & $\log L_K$& $\log M_{orb}$& $M_{orb}/L_K$\\
\hline
	      &  Mpc       &    & mag    &        kpc  &   km/s         &             &       &\\
\hline
 IC 342        &3.28        &8    &3.0           &319     &76           &10.60     &12.28      &48\\
 M 101         &6.95        &9    &3.2           &204     &64           &10.79     &11.93       &14\\
 NGC 6946      &7.73       &8     &5.8           &323     &68           &10.99     &12.05        &11\\
 NGC 628      &10.19       &9     &4.6           &309     &68           &10.60     &12.11        &32\\
 NGC 3184     &11.12       &5     &3.3          &454     &49           &10.52     &11.87        &22\\
\hline
 NGC 253       &3.94       &7     &2.0          &500     &60           &11.04     &12.18        &14\\
 NGC 5236      &4.92      &10     &4.4          &294     &61            &10.86    &12.03       &15\\
\hline
 Mean               &-     &8     &3.8         &343     &65             &10.77    &12.06       &22\\ \hline
\end{tabular}
\end{table}
   Table 3 contains the basic parameters  of the nearby groups. As seen from the data, even the brightest satellites appear to be fainter of the central galaxy at 3--5 magnitudes. Thus, the application of the model of test particles moving around central massive body is quite correct in evaluating the total mass of the central galaxy.
   
   In the Local Volume there are two another groups, NGC~253 and NGC~5236, where the dominated member is a spiral bulge-less galaxy but oriented arbitrary. We give their data in the bottom of Table 3. The  K-band luminosities and orbital mass estimates for these nearby groups are rather similar to ones from Table 1 obtained for more distant systems. Particularly, the mean total mass-to-luminosity ratio for the Local Volume bulge-less galaxies, $\langle M_{orb} /L_K\rangle = 22\pm5$, is practically the same as the value ($20\pm3$) derived for the remote Sc-Sd galaxies.
     
     It should be noted that according to the data by Karachentsev \& Kudrya (2014), the Local Volume contains five groups: M~31, M~81, NGC~4258, NGC~4736, and NGC~3627, where the bulgy Sab--Sbc galaxies dominate, and also three groups: NGC~5128, NGC~4594, NGC~3115, where the objects are grouping around the E, S0, Sa galaxies. These groups are characterized of the mean values of total mass-to-luminosity ratio $\langle M_{orb} /L_K\rangle = 41\pm9$ and $\langle M_{orb} /L_K\rangle = 88\pm30$, respectively. In spite of poor statistics, these results can indicate that the dark mass-to-luminous mass ratio grows when  the fraction of  galaxy spheroidal stellar sub-system increases.  

    \section{Concluding remarks.}
    As we noticed above, the thin spiral bulge-less galaxies reside in the low density regions, i.e. have a poor environment. The rare satellites of the Sc--Scd--Sd galaxies are, as a rule, dwarf galaxies containing  gas and young stars. Our statistics of satellites around 220 face-on  bulge-less galaxies shows that they have the average projected separation of  $\langle R_p\rangle =130$ kpc and mean-square velocity difference of  96 km/s. The estimate of the total mass of bulge-less galaxies via orbital motions of their satellites yields the value of $\langle M_{orb}\rangle =12.14$ dex  and  $\langle M_{orb}/L_K\rangle = 20\pm3$.  A similar value, $\langle M_{orb}/L_K\rangle = 22\pm5$,  is obtained by us for the Local Volume Sc--Sd galaxies, where the separation  of physical satellites and field galaxies is more reliable due to abundance of data on the galaxy distances.  For comparison, the Sab--Sbc spirals in the Local Volume with much more developed bulges have the ratio $\langle M_{orb}/L_K\rangle = 41\pm9$, and a few in number galaxies of E, S0, Sa types are characterised by the ratio of $\langle M_{orb}/L_K\rangle = 88\pm30$.  It seems unlikely that this difference would be caused by different manner of transition from the B-luminosity of a galaxy to its K-luminosity, or the different nature of the orbital motions of the satellites. Obviously, the conclusion
about the reduced amount of dark matter per unit of K-band luminosity in the Sc--Sd  galaxies needs confirmation on a richer observational material. It is quite possible that the presence of  ''scraggy'' halos in spiral bulge-less galaxies  will be explained within the framework of the standard paradigm of the hierarchy galaxy clustering.

{\bf Acknowledgements.}

 We thank the referee for his/her valuable comments. The work is supported by the Russian Science Foundation grant 19-12-00145.
The paper makes use of the data from NASA Extragalactic Database and from the HyperLEDA.

{\bf REFERENCES}

 Banerjee A., \& Jog C.J., 2013, MNRAS, 431, 582

  Barber C., Starkenburg E., Navarro J.F.,  et al. 2014, MNRAS, 437, 959

  de Vaucouleurs, G., de Vaucouleurs, A., and Corwin, H. G. 1976, Second
   Reference Catalogue of Bright Galaxies, Austin: University of Texas Press (RC2)
   
 Kaisina E.I., Makarov D.I., Karachentsev I.D., Kaisin S.S., 2012, AstBu, 67, 115

Karachentsev I.D. \&  Karachentseva V.E., 2019, MNRAS, 485, 1477

Karachentsev I.D.,   Karachentseva V.E., Kudrya Y.N., 2016, AstBu, 71, 129

Karachentsev I.D. \& Kudrya Y.N., 2014, AJ, 148, 50

Karachentsev I.D., Makarov D., Kaisina E., 2013, AJ, 145, 101 (UNGC)

Karachentsev I.D., Karachentseva V.E., Kudrya Y.N., et al., 1999, AstBu, 47, 5 (RFGC)

Karachentseva V.E., Kudrya Y.N., Karachentsev I.D., Makarov D.I., Melnyk O.V., 2016, AstBu, 71, 1 (UF)

Kormendy J., Drory N., Bender R., Cornell M., 2010, ApJ, 723, 54

Kormendy, J. \& Kennicutt, R.C., 2004, ARA\&A, 42, 603

Makarov D., Prugniel P., Terekhova N., et al. 2014, A\&A, 570A, 13 (HyperLEDA)

Makarov D.I., Burenkov A.N., Tyurina N.V., 2001, AstL, 27, 213

Melnyk O.V., Karachentseva V.E., Karachentsev I.D., 2017, AstBu, 72, 1

Uson J.M., Matthews L.D., 2003, AJ, 125, 2455


\end{document}